\documentclass[twocolumn,showpacs,preprintnumbers,amsmath,amssymb]{revtex4}
\usepackage{graphicx}
\usepackage{dcolumn}
\usepackage{bm}
\def\jpsi{{J/\psi}}
\def\uo{{\Upsilon^{(8)}}}
\def\jpsio{{\Upsilon\bigl[\bigl.^1\hspace{-1mm}S^{(8)}_0,\bigl.^3\hspace{-1mm}S^{(8)}_1\bigr]}}
\def\jpsioa{{\Upsilon\bigl[^1\hspace{-1mm}S^{(8)}_0\bigr]}}
\def\jpsiob{{\Upsilon\bigl[^3\hspace{-1mm}S^{(8)}_1\bigr]}}

\def\sa{{\bigl.^1\hspace{-1mm}S^{(8)}_0}}
\def\sb{{\bigl.^3\hspace{-1mm}S^{(8)}_1}}
\def\pj{{\bigl.^3\hspace{-1mm}P^{(8)}_J}}
\def\mop{{\langle\mathcal{O}^H_n\rangle}}
\def\mopa{{\langle\mathcal{O}^\Upsilon_8(\bigl.^1\hspace{-1mm}S_0)\rangle}}
\def\mopb{{\langle\mathcal{O}^\Upsilon_8(\bigl.^3\hspace{-1mm}S_1)\rangle}}

\def\ME#1#2#3{{\langle\mathcal{O}^{#1}_{#2}#3\rangle}}
\def\OP#1#2#3{{\bigl.^{#1}\hspace{-1mm}{#2}_{#3}}}

\def\be{\begin{equation}}
\def\ee{\end{equation}}
\def\bea{\begin{eqnarray}}
\def\eea{\end{eqnarray}}
\def\NO{\nonumber}
\def\gev{\mathrm{~GeV}}

\def\dfrac{\displaystyle\frac}
\def\dg{\sp\dagger}
\def\md{\mathrm{d}}
\def\li{\mathrm{Li_2}}
\def\co{{\cal O}}
\def\a{\alpha}
\def\b{\beta}

\def\d{\delta}
\def\e{\epsilon}
\def\F{\Phi}
\def\g{\gamma}
\def\G{\Gamma}
\def\m{\mu}

\def\s{\sigma}
\def\q{\theta}
\def\p{\pi}

\begin{document}


\title{QCD corrections to $\Upsilon$ production via color-octet states at the Tevatron and LHC}

\author{Bin Gong$^{1,2,3}$, Jian-Xiong Wang$^{1,3}$ and Hong-Fei Zhang$^{1,3}$}%
\affiliation{
Institute of High Energy Physics, CAS, P.O. Box 918(4), Beijing, 100049, China. \\
Institute of Theoretical Physics, CAS, P.O. Box 2735, Beijing, 100190, China. \\
Theoretical Physics Center for Science Facilities, CAS, Beijing, 100049, China.
}%

\date{\today}

\begin{abstract}
The NLO QCD corrections to $\Upsilon$ production via S-wave color-octet states $\jpsio$ at the
Tevatron and LHC is calculated. The K factors of total cross section (ratio of NLO to LO) are 1.313 and 1.379 for $\jpsioa$ and $\jpsiob$ at the Tevatron, while at the LHC they are 1.044 and 1.182, respectively. By fitting the experimental data from the D0, the matrix elements for S-wave color-octet states are obtained. And new predictions for $\Upsilon$ production are presented. The prediction for the polarization of inclusive $\Upsilon$ contains large uncertainty rising from the polarization of $\Upsilon$ from feed-down of $\chi_b$. To further clarify the situation, new measurements on the production and polarization for direct $\Upsilon$ are expected. 
\end{abstract}

\pacs{12.38.Bx, 13.25.Gv, 13.60.Le}
\maketitle
\section{Introduction}
For heavy quarkonium production and decay, a naive perturbative QCD and nonrelativistic factorization treatment is applied straightforwardly. It is called color-singlet mechanism (CSM). To describe the huge discrepancy of the high-$p_t$ $J/\psi$ production between the theoretical prediction based on CSM and the experimental measurement at Tevatron, a color-octet mechanism~\cite{Braaten:1994vv} was proposed based on the non-relativistic QCD (NRQCD)~\cite{Bodwin:1994jh}. In the application, $\jpsi$ or $\Upsilon$ related productions or decays are very good places for two reasons, theoretically charm and bottom quarks are thought to be heavy enough, so that charmonium and bottomonium can be treated within the NRQCD framework, experimentally there is a very clear signal to detect $\jpsi$ and $\Upsilon$. The key point is that the color-octet mechanism depends on nonperturbative universal NRQCD matrix elements, which is obtained by fitting the data. Therefore various efforts have been made to confirm this mechanism, or to fix the magnitudes of the universal NRQCD matrix elements. Although it seems to show qualitative agreements with experimental data, there are certain difficulties. A
review of the situation could be found in Refs.~\cite{Kramer:2001hh,Lansberg:2006dh}.

To explain the experimental 
measurements~\cite{Abe:2001za,Aubert:2005tj} of $J/\psi$ production at the B factories, a
series of calculations~\cite{Zhang:2005cha,Gong:2009ng} in the CSM reveal that the next-to-leading order (NLO) QCD corrections give 
the main contribution to the related processes. Together with the relativistic correction~\cite{Bodwin:2006ke}, 
it seems that most experimental data for $J/\psi$ production at the B factories could be understood.
Recent studies show that the NLO QCD correction also plays an 
important role in $J/\psi$ production at RHIC~\cite{Brodsky:2009cf} and the 
hadroproduction of $\chi_c$~\cite{Ma:2010vd}. 
For the $J/\psi$ photoproduction, the $p_t$ and $z$ distributions can be described by the NLO calculations 
in CSM~\cite{Kramer:1995nb} by choosing a small renormalization scale, but recent NLO calculations 
in CSM \cite{Artoisenet:2009xh} show that the $p_t$ distributions of the production and polarization for 
$J/\psi$ can not be well described when choosing a proper renormalization scale. 
Although the complete calculation at NLO in COM~\cite{Butenschoen:2009zy} can account for the 
experimental measurements on the $p_t$ distribution, it cannot extend to $J/\psi$ polarization case.
To further study the heavy quarkonium production mechanism, there are
many other efforts performed, such as NLO QCD correction to $J/\psi$ production associated with 
photon~\cite{Li:2008ym}, QED contributions in $J/\psi$ hadroproduction~\cite{He:2009cq},  
inclusive $J/\psi$ production from $\Upsilon$ decay~\cite{He:2009by}, 
double heavy quarkonium hadronproduction~\cite{Li:2009ug}, and NLO QCD correction to $J/\psi$ production from $Z$ decay~\cite{Li:2010xu}.

For the polarized heavy quarkonium hadroproduction, the leading order (LO) NRQCD prediction gives a sizable transverse polarization for $\jpsi$ production at high $p_t$ at Tevatron while the experimental measurement~\cite{Abulencia:2007us} gives slight longitudinal polarized result. The discrepancy was also found in $\Upsilon$ production. In a recent paper~\cite{Abazov:2008za}, the measurement on polarization of $\Upsilon$ production at Tevatron is presented and the NRQCD prediction~\cite{Braaten:2000gw} is not coincide with it. Within the NRQCD framework, higher order correction is thought to be an important way towards the solution of such puzzles. Recently, NLO QCD corrections to $\jpsi$ and $\Upsilon$ hadroproduction have been calculated~\cite{Campbell:2007ws,Qiao:2003ba,Artoisenet:2007xi,Gong:2008sn,Gong:2008hk}, and the results show that the NLO QCD corrections give significant enhancement to both total cross section and momentum distribution for the color-singlet channel. This would reduce the contribution of color-octet channel in the production. Also, it is found in Ref.~\cite{Gong:2008sn} that the polarizations for $\jpsi$ and $\Upsilon$ hadroproduction via color-singlet channel would change drastically from transverse polarization dominant at LO into longitudinal polarization dominant in the whole range of the
transverse momentum $p_t$ at NLO. It seems that these results open a door to the solution of the problem. But things are not always going as expected. The NLO QCD corrections to the $\jpsi$ production via S-wave color-octet states were studied in our previous work~\cite{Gong:2008ft}. It was found that the effect of NLO QCD correction is small and the discrepancy holds on. For the color-singlet part,  the partial next-to-next-to-leading order (NNLO) calculations for $\Upsilon$ and $J/\psi$ hadroproduction show that the uncertainty from higher order QCD 
correction~\cite{Artoisenet:2008fc} is quite large, therefore no definite conclusion can be made.
As we know, the contribution from the color-octet states is smaller in the $\Upsilon$ production than that in $\jpsi$ production, thus things may be different. In this paper, we present our calculation on NLO QCD corrections to $\Upsilon$ hadroproduction via S-wave color-octet states. New matrix elements are fitted and new prediction for the polarization status is presented.

This paper is organized as follows. In Sec. II, we give the LO cross section for the process.
The calculation of NLO QCD corrections are described in Sec. III. In Sec. IV, we present the formula in final integration to obtain the transverse momentum distribution of $\Upsilon$ production. Sec. V. is devoted to the description about the calculation of $\Upsilon$ polarization. The treatment of $\jpsi$ is discussed in Sec. VI. The numerical results are presented in Sec. VII, while the summary and discussion are given in Sec. VIII. In the Appendix, several details of the calculation are presented.

\section{The LO cross section}
\begin{figure*}
\center{
\includegraphics*[scale=0.8]{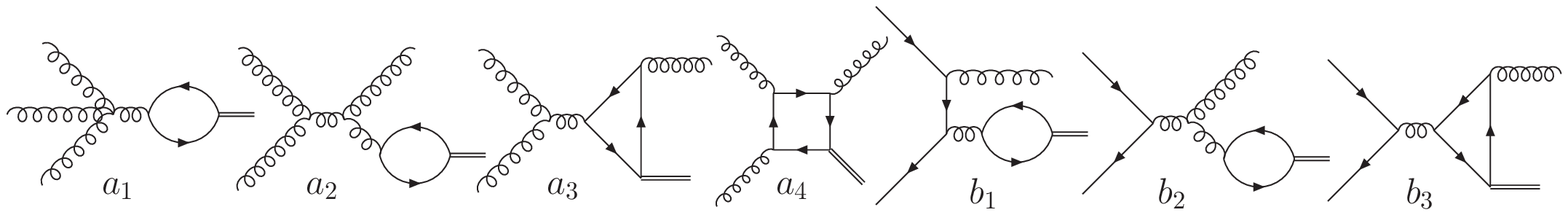}
\caption {\label{fig:lo}Typical Feynman diagrams for LO processes. $a$) Feynman diagrams for process (\ref{prs:lo_ggg}); $b$) Feynman diagrams for processes (\ref{prs:lo_gqq}) and (\ref{prs:lo_qqg}). Diagrams in groups $(a_1),~(a_2),~(b_1)$ and $(b_2)$ are absent for the $\sa$ state. }}
\end{figure*}
According to the NRQCD factorization formalism,
the inclusive cross section for direct $\Upsilon$ production in hadron-hadron collision is expressed as
\bea
\s[pp\rightarrow \Upsilon+X]&=\sum\limits_{i,j,k,n}\int \mathrm{d}x_1\mathrm{d}x_2 G_{i/p}G_{j/p} \NO\\
&\times\hat{\s}[i+j\rightarrow (b\bar{b})_n +k]\mop, \eea where $p$
is either a proton or an anti-proton, the indices $i, j,k$ run over all
the partonic species and $n$ denotes the color, spin and angular
momentum states of the intermediate $b\bar{b}$ pair. The
short-distance contribution  $\hat{\s}$ can be perturbatively
calculated order by order in  $\a_s$. The hadronic matrix elements
$\mop$ are related to the hadronization from the state
$(b\bar{b})_n$ into $\Upsilon$ which are fully governed by the
non-perturbative QCD effects. In the following, $\hat{\s}$ represents the corresponding partonic cross section.

At LO, there are three partonic processes:
\begin{align}
&&g(p_1)+ g(p_2) \rightarrow \jpsio(p_3) + g(p_4)  \tag{L1},\label{prs:lo_ggg} \\
&&g(p_1)+ q(p_2) \rightarrow \jpsio(p_3) + q(p_4)   ,\label{prs:lo_gqq} \tag{L2} \\
&&q(p_1)+ \overline{q}(p_2) \rightarrow \jpsio(p_3) + g(p_4) .\label{prs:lo_qqg} \tag{L3}
\end{align}
where $q$ represents a sum over all possible light quarks or
anti-quarks: $u,~d,~s,~c,~\overline{u},~\overline{d},~\overline{s}$ and $\bar{c}$. In our calculation of $\Upsilon$ production, we take charm quark as light quark as an approximation. Typical Feynman diagrams for these three processes are shown in Fig.~\ref{fig:lo}. And the partonic differential cross sections in $n=4-2\e$ dimension for LO processes can be obtained as
\begin{widetext}
\bea
\dfrac{\mathrm{d}\hat{\s}^{B}(q\overline{q}\rightarrow \jpsiob g)}
{\mathrm{d}\hat{t}}&=&
\dfrac{\pi^2\a_s^3\mopb [(\hat{t}-1)^2+(\hat{u}-1)^2][4\hat{t}^2-\hat{t}\hat{u}+4\hat{u}^2]}
{324m_b^5\hat{s}^2(\hat{s}-1)^2\hat{t}\hat{u}} + \co(\e), \NO\\
\dfrac{\mathrm{d}\hat{\s}^{B}(gq\rightarrow \jpsiob q)}
{\mathrm{d}\hat{t}}&=&
\dfrac{-\pi^2\a_s^3\mopb [(\hat{s}-1)^2+(\hat{u}-1)^2][4\hat{s}^2-\hat{s}\hat{u}+4\hat{u}^2]}
{864m_b^5\hat{s}^3(\hat{t}-1)^2\hat{u}} + \co(\e), \NO\\
\dfrac{\mathrm{d}\hat{\s}^{B}(gg\rightarrow \jpsiob g)}
{\mathrm{d}\hat{t}}&=&
\dfrac{\pi^2\a_s^3\mopb [(\hat{s}^2-1)^2+(\hat{t}^2-1)^2+(\hat{u}^2-1)^2-6\hat{s}\hat{t}\hat{u}-2] [19-27(\hat{s}\hat{t}+\hat{t}\hat{u}+\hat{u}\hat{s})]}
{1152m_b^5\hat{s}^2(\hat{t}-1)^2(\hat{u}-1)^2(\hat{s}-1)^2} + \co(\e), \NO\\
\dfrac{\mathrm{d}\hat{\s}^{B}(q\overline{q}\rightarrow \jpsioa g)}
{\mathrm{d}\hat{t}}&=&
\dfrac{5\pi^2\a_s^3\mopa [\hat{t}^2+\hat{u}^2]}
{216m_b^5\hat{s}^3(\hat{s}-1)^2} + \co(\e), \NO\\
\dfrac{\mathrm{d}\hat{\s}^{B}(gq\rightarrow \jpsioa q)}
{\mathrm{d}\hat{t}}&=&
\dfrac{-5\pi^2\a_s^3\mopa [\hat{s}^2+\hat{u}^2]}
{576m_b^5\hat{s}^2\hat{t}(\hat{t}-1)^2} + \co(\e), \NO\\
\dfrac{\mathrm{d}\hat{\s}^{B}(gg\rightarrow \jpsioa g)}
{\mathrm{d}\hat{t}}&=&
\dfrac{5\pi^2\a_s^3\mopa [\hat{s}^2\hat{t}^2+\hat{s}^2\hat{u}^2+\hat{t}^2\hat{u}^2+\hat{s}\hat{t}\hat{u}] [\hat{s}^4+\hat{t}^4+\hat{u}^4+1]}
{256m_b^5\hat{s}^3\hat{t}\hat{u}(\hat{t}-1)^2(\hat{u}-1)^2(\hat{s}-1)^2} + \co(\e), \NO\\
\eea
\end{widetext}
by introducing three dimensionless kinematic variables:
\be
\hat{s}=\dfrac{(p_1+p_2)^2}{4m_b^2},\quad \hat{t}=\dfrac{(p_1-p_3)^2}{4m_b^2}, \quad \hat{u}=\dfrac{(p_1-p_4)^2}{4m_b^2},
\ee
and the reasonable approximation $M_{\Upsilon}=2m_b$ is taken. Our LO results are consistent with those in Ref.~\cite{Cho:1995ce}.
The LO total cross section is obtained by convoluting the partonic cross section
with the parton distribution function (PDF) in the proton:
\bea
\s^B[pp\rightarrow \uo+X]&=&\sum\limits_{i,j,k}\int  \hat{\s}^B[i+j\rightarrow \uo +k] \\
&\times & G_{i/p}(x_1,\m_f)G_{j/p}(x_2,\m_f)\mathrm{d}x_1\mathrm{d}x_2,  \NO
\eea
where $\uo$ denotes certain color-octet $\jpsioa$ or $\jpsiob$, $\mu_f$ is the factorization scale.

\section{The NLO cross section}
The NLO contributions can be written as a sum of two
parts: first is the virtual corrections which arise from loop
diagrams, the other is the real corrections caused by radiation of a
real gluon, or a gluon splitting into a light quark-antiquark pair,
or a light (anti)quark splitting into a light (anti) quark and a
gluon.
\subsection{Virtual corrections}
There exist ultraviolet (UV), infrared (IR) and Coulomb singularities in the calculation of the
virtual corrections. UV divergences from self-energy and triangle
diagrams are canceled by introducing renormalization. Here we adopt
the renormalization scheme used in Ref.~\cite{Klasen:2004tz}. The
renormalization constants $Z_m$, $Z_2$, $Z_{2l}$ and $Z_3$ which
correspond to bottom quark mass $m_b$, bottom-field $\psi_b$, light
quark field $\psi_q$ and gluon field $A^a_{\mu}$ are defined in the
on-mass-shell (OS) scheme while $Z_g$ for the QCD gauge coupling constant
$\alpha_s$ is defined in the
modified-minimal-subtraction ($\overline{\mathrm{MS}}$) scheme: \bea
\delta Z_m^{OS}&=&-3C_F\dfrac{\alpha_s}{4\pi}\left[\dfrac{1}{\e_{UV}} -\gamma_E +\ln\dfrac{4\pi \mu_r^2}{m_b^2} +\frac{4}{3}\right] ,\NO \\
\delta Z_2^{OS}&=&-C_F\dfrac{\alpha_s}{4\pi}\left[\dfrac{1}{\e_{UV}} +\dfrac{2}{\e_{IR}} -3\gamma_E +3\ln\dfrac{4\pi
\mu_r^2}{m_b^2} +4 \right] ,\NO \\
\delta Z_{2l}^{OS}&=&-C_F\dfrac{\alpha_s}{4\pi}\left[
\dfrac{1}{\e_{UV}} -\dfrac{1}{\e_{IR}}
\right] ,\NO \\
\delta Z_3^{OS}&=&\dfrac{\alpha_s}{4\pi}\left[(\beta_0-2C_A)\left(\dfrac{1}{\e_{UV}} -\dfrac{1}{\e_{IR}}\right)
\right] , \\
\delta
Z_g^{\overline{\mathrm{MS}}}&=&-\dfrac{\beta_0}{2}\dfrac{\alpha_s}{4\pi}\left[\dfrac{1}{\e_{UV}}
-\gamma_E +\ln(4\pi)\right] \NO, \eea where $\g_E$ is the Euler's
constant, $\b_0=\frac{11}{3}C_A-\frac{4}{3}T_Fn_f$ is the one-loop
coefficient of the QCD beta function and $n_f$ is the number of
active quark flavors. We have four light quarks $u$, $d$, $s$ and $c$ in our calculation, so $n_f$=4. The
color factors are given by $T_F=1/2, C_F=4/3, C_A=3$
and $\mu_r$ is the renormalization scale.
\begin{figure*}
\center{
\includegraphics*[scale=0.7]{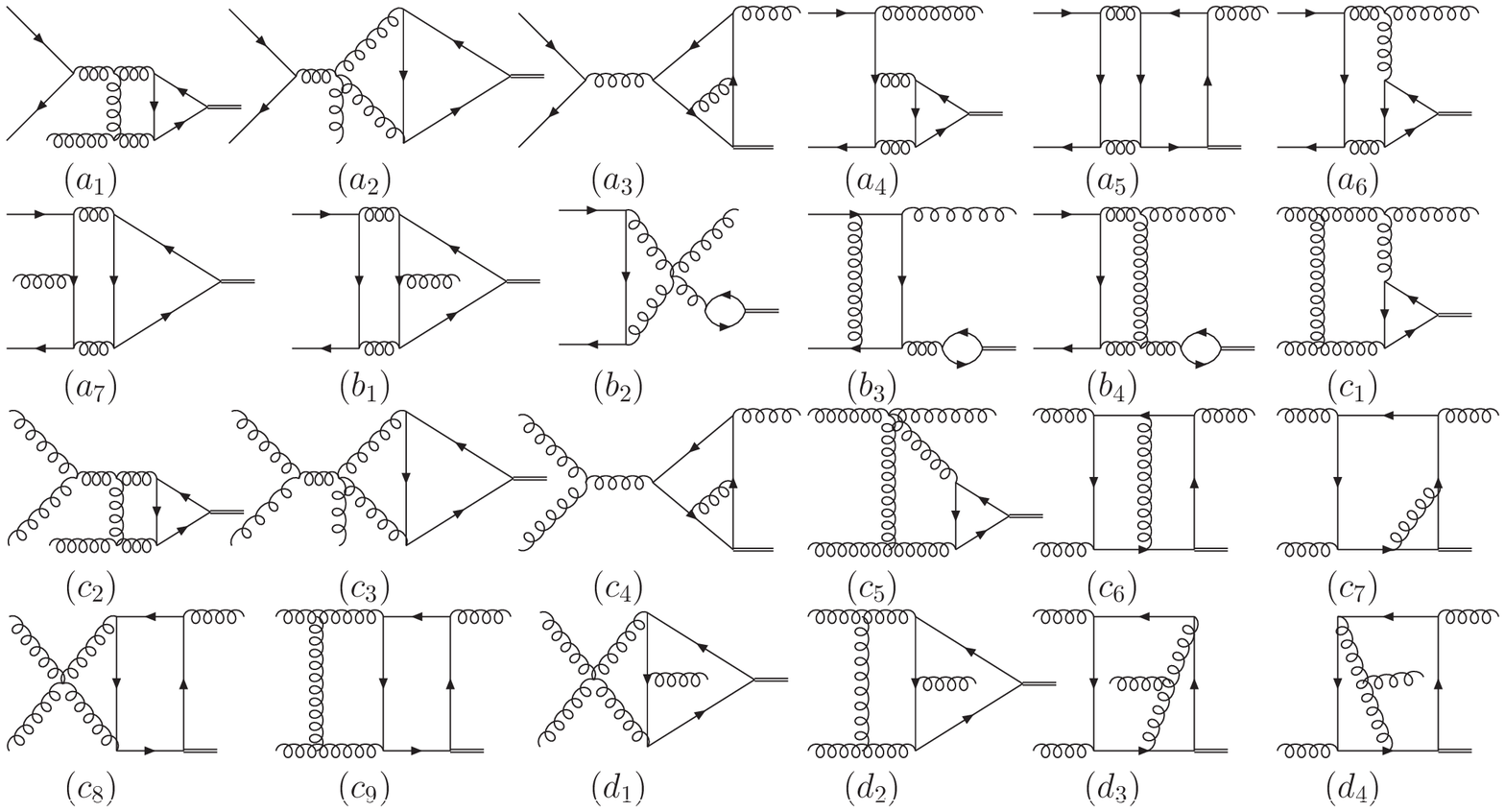}
\caption {\label{fig:nlo}Typical one-loop diagrams. $a$) Feynman diagrams for $gq\rightarrow \jpsioa q$ and $q\bar{q}\rightarrow \jpsioa g$; $a+b$) Feynman diagrams for $gq\rightarrow \jpsiob q$ and $q\bar{q}\rightarrow \jpsiob g$; $c$) Feynman diagrams for $gg\rightarrow \jpsioa g$; $c+d$) Feynman diagrams for $gg\rightarrow \jpsiob g$. Counter-term diagrams, together with corresponding loop diagrams, are not shown here. }}
\end{figure*}

There are 267 (for the $\sa$ state) and 413 (for the $\sb$ state) NLO diagrams for process (\ref{prs:lo_ggg}), including counter-term
diagrams, while for both processes (\ref{prs:lo_gqq}) and
(\ref{prs:lo_qqg}), there are 49 (for the $\sa$ state) and 111 (for
the $\sb$ state) NLO diagrams. Part of the Feynman diagrams for these processes are shown in Fig.~\ref{fig:nlo}. The diagrams in which a virtual
gluon line connects the quark pair possess  Coulomb singularities,
which can be isolated and attributed into  renormalization of the
$b\bar{b}$ wave function.

For each process, by summing over contributions from all diagrams,
the virtual correction to the differential cross section can be
expressed as 
\be 
\dfrac{\mathrm{d}\hat{\s}^{V}_{[{\mathrm{L}_i}]}}{\mathrm{d}t}
\propto 2\mathrm{Re}\left(M^B_{[\mathrm{L}_i]}M^{V^*}_{[\mathrm{L}_i]}\right), 
\label{eqn:virtual_sme}
\ee where $M^B_{[\mathrm{L}_i]}$ is the
amplitude of process ($\mathrm{L}_i$) at LO, and $M^V_{[\mathrm{L}_i]}$ is the renormalized
amplitude of corresponding process at NLO. $M^V_{[\mathrm{L}_i]}$ is UV and Coulomb
finite, but it still contains IR divergences. 
And the total cross section of virtual contribution could be written as
\bea
\s^V[pp\rightarrow \uo+X]&=&\sum\limits_{i,j,k}\int  \hat{\s}^V[i+j\rightarrow \uo +k] \\
&\times & G_{i/p}(x_1,\m_f)G_{j/p}(x_2,\m_f)\mathrm{d}x_1\mathrm{d}x_2,  \NO
\eea
\subsection{Real corrections}
\begin{figure*}
\center{
\includegraphics*[scale=0.8]{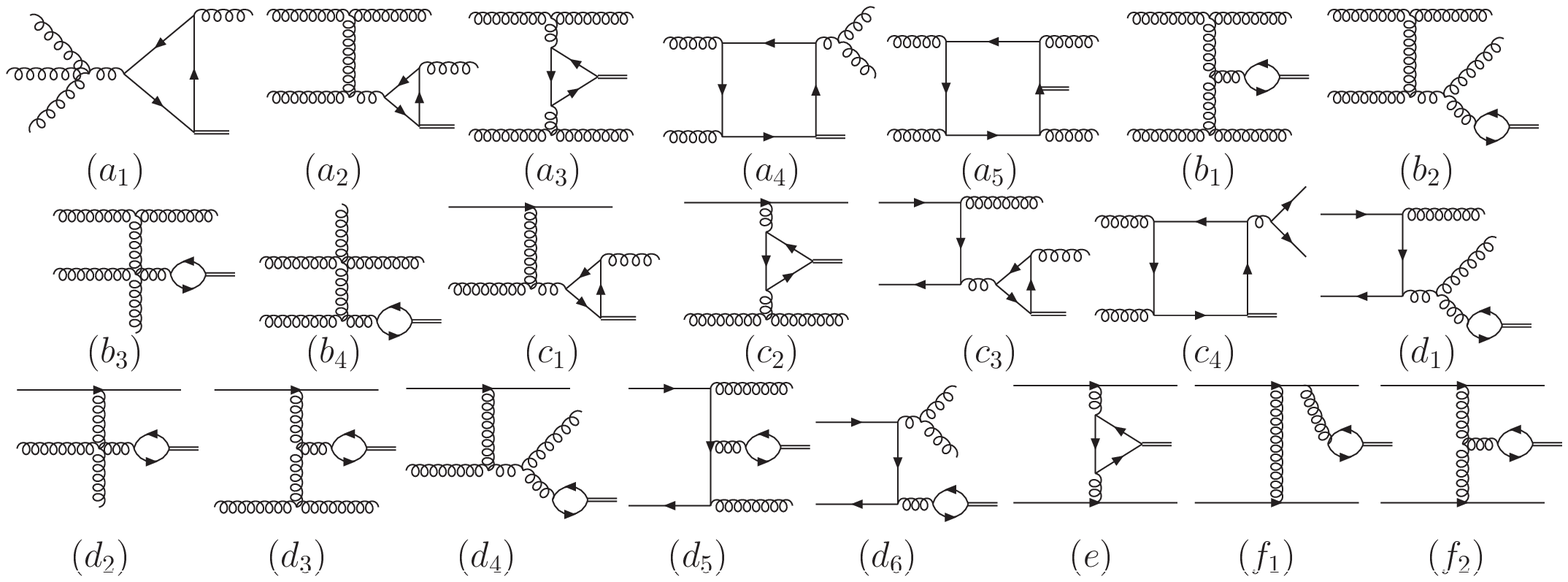}
\caption {\label{fig:real}Feynman diagrams for real correction processes. $a$) R1 ($\jpsioa$); $a+b$) R1 ($\jpsiob$); $c$) R2$\sim$R4 ($\jpsioa$); $c+d$)  R2$\sim$R4 ($\jpsiob$); $e$) R5$\sim$R8 ($\jpsioa$); $e+f$) R5$\sim$R8 ($\jpsiob$). R1 ($\jpsioa$) denotes process $gg\rightarrow \jpsioa gg$, R1 ($\jpsiob$) denotes process $gg\rightarrow \jpsiob gg$, and so on.}}
\end{figure*}

There are eight processes involved in the real corrections: 
\begin{align}
gg&\rightarrow  \jpsio gg, \tag{R1} \label{prs:gggg} \\
gq&\rightarrow\jpsio gq,\tag{R2}\label{prs:qggq} \\
q\overline{q}&\rightarrow \jpsio gg ,\tag{R3}\label{prs:qqgg} \\
gg&\rightarrow \jpsio q\overline{q},\tag{R4}\label{prs:ggqq} \\
q\overline{q}&\rightarrow \jpsio q\overline{q} ,\tag{R5}\label{prs:qqqq0} \\
q\overline{q}&\rightarrow \jpsio q'\overline{q}' ,\tag{R6}\label{prs:qqqq1} \\
qq&\rightarrow \jpsio qq ,\tag{R7}\label{prs:qqqq2} \\
qq'&\rightarrow \jpsio qq' ,\tag{R8}\label{prs:qqqq3} 
\end{align}
where $q, q'$ denote light quarks (anti-quarks) with different
flavors. Feynman diagrams for these processes are shown in Fig.~\ref{fig:real}.
We have neglected the contributions from the another two processes, $gg\rightarrow \uo b\bar{b}$ and $q\bar{q}\rightarrow \uo b\bar{b}$, which are IR finite and small. 
Phase space integrations of above eight processes generate IR
singularities, which are either soft or collinear and can be
conveniently isolated by slicing the phase space into different
regions. We use the two-cutoff phase space slicing method
\cite{Harris:2001sx}, which introduces two small cutoffs to
decompose the phase space into three parts. Then the real cross
section can be written as \be
\s^R=\s^S+\s^{HC}+\s^{H\overline{C}} .\ee 
It is easy
to observe that different parts of IR singularities from one real 
process may be factorized and each part should be added into the
cross sections of different LO processes. This is the reason why we
have to calculate the NLO corrections to the three LO processes
together.
\subsubsection{soft}
Soft singularities arise from real gluon emission. Thus only real processes (\ref{prs:gggg}), (\ref{prs:qggq}) and (\ref{prs:qqgg}) contain soft singularities, corresponding to the three LO processes. One should notice that, unlike 
color-singlet case, the soft singularities caused by emitting a soft gluon from the quark pair in the S-wave color octet exists and we find that the factorized matrix element is the same as the case of emitting a soft gluon from a gluon. 

Suppose $p_5$ is the momentum of the emitted gluon. If we define the
Mandelstam invariants as $s_{ij}=(p_i+p_j)^2$ and $t_{ij}=(p_i-p_j)^2$,
the soft region is defined in term of the energy of $p_5$ in the $p_1+p_2$ rest frame
by $0 \leq E_5 \leq \d_s \sqrt{s_{12}}/2$. 
For each of the three real processes, 
$\hat{\s}^S$ from the soft regions is calculated analytically under the soft approximation.

Following the similar factorization procedure as applied in the calculation of color-singlet case~\cite{Gong:2008hk}, the matrix elements for a certain real process ($\mathrm{R}_i$) in the soft region can be written as 
\be
|M_{[\mathrm{R}_i]}|^2|_{\rm soft} \simeq -4\pi\a_s \mu_r^{2\e} \sum_{j,k=1}^4
\frac{-p_j \cdot p_k}{(p_j \cdot p_5)(p_k\cdot p_5)} M_{[\mathrm{L}_i]}^{jk} \, ,
\label{eqn:sme_soft}
\ee
with
\be
M_{[\mathrm{L}_i]}^{jk} =
\left[{\bf T}^a(j) {\bf M}^{b_1\cdots b_{j^\prime}\cdots b_4}_{[\mathrm{L}_i]}\right]^{\dg} \left[{\bf T}^a(k)
{\bf M}^{b_1\cdots b_{k^\prime}\cdots b_4}_{[\mathrm{L}_i]} \right] \,
\label{eqn:me0_cc}
\ee
where ${\bf M}^{b_1\cdots b_4}_{[\mathrm{L}_i]}$ is the color connected Born matrix element for LO processes ($\mathrm{L}_i)$. If the emitting parton $j$ is an initial state quark or a final state antiquark, ${\bf T}^a(j)=T^a_{b_{j^\prime}b_j}$. For an initial state antiquark or a final state quark ${\bf T}^a(j)=-T^a_{b_jb_{j^\prime}}$. If the emitting parton $j$ is a gluon or the color-octet state, ${\bf T}^a(j)=if_{ab_jb_{j^\prime}}$.
And the corresponding parton level differential cross section can be expressed as
\be
d\hat{\s}^S_{[\mathrm{R}_i]} =
\left[ \frac{\a_s}{2\p} \frac{\G(1-\e)}{\G(1-2\e)} \left( \frac{4\p\m_r^2}{s_{12}} \right)^\e \right] 
\sum_{j,k=1}^4 d\hat{\s}^{jk}_{[\mathrm{L}_i]} I^{jk} \, ,
\label{eqn:soft_final}
\ee
with
\be
d\hat{\s}_{[\mathrm{L}_i]}^{jk} = \frac{1}{2\F} \overline{\sum} M_{[\mathrm{L}_i]}^{jk} d\G_2 \, .
\label{eqn:soft_lo_c_o}
\ee
The factor $I^{jk}$ is universal for all three real processes, and is given in Appendix.~(\ref{chapter:I_jk}). Sometimes 
$d\hat{\s}_{[\mathrm{L}_i]}^{jk}$ may be written in a more compact form as
\be 
d\hat{\s}_{[\mathrm{L}_i]}^{jk}=C^{jk}_{[\mathrm{L}_i]} d\hat{\s}^B_{[\mathrm{L}_i]},
\label{eqn:soft_lo_c}
\ee
where $C^{jk}_{[\mathrm{L}_i]}$ is a constant. This is always true if the LO process ($\mathrm{L}_i$) contain only one independent color factor in the matrix element. But for processes with two or more than two independent color factors, there seems no sure reason for it to be or not to be true. Of course, no matter Eq.~(\ref{eqn:soft_lo_c}) is true or not, we can always obtain $d\hat{\s}_{[\mathrm{L}_i]}^{jk}$ through Eq.~(\ref{eqn:soft_lo_c_o}). Most processes involved in this calculation have more than one independent color factors, and they are listed in Appendix.~(\ref{chapter:color_factors_lo}).
\subsubsection{hard collinear}
The hard collinear regions of the phase space are those where any invariant
($s_{ij}$ or $t_{ij}$) becomes smaller in magnitude than $\d_c s_{12}$.
It is treated according to whether the singularities are from initial or final
state emitting or splitting in the origin. 
\paragraph{final state collinear}
For real processes (R1) $\sim$ (R6), which contain final state collinear singularities, the final state collinear
region is defined by $0 \le s_{45} \le \d_c s_{12}$. Again following the similar factorization procedure described in 
Ref~\cite{Harris:2001sx}, the parton level cross section in the hard final state collinear region can be expressed as
\be 
\hat{\s}^{HC}_{f}[\mathrm{R}_i]=\hat{\s}^B[\mathrm{L}_i^\prime]\left[ \frac{\a_s}{2\pi} \frac{\Gamma(1-\e)}{\Gamma(1-2\e)} \left( \frac{4\pi\mu_r^2}{s_{12}} \right)^\e \right]A^{HC}_{i}.
\ee
For a certain real process ($\mathrm{R}_i$), ($\mathrm{L}_i^\prime$) is the corresponding LO process it factorizes into. And the coefficient $A^{HC}_{i}$ are listed in Table.~\ref{table:final_coll}, with 
\bea
A_1^{g \rightarrow gg} &=& N \left( 11/6 + 2 \ln\d'_s \right) \nonumber\\
A_0^{g \rightarrow gg} &=& N \left[ 67/18 - \pi^2/3 - \ln^2\d'_s
    - \ln\d_c \left( 11/6 + 2 \ln\d'_s \right) \right] \nonumber\\
A_1^{q \rightarrow qg} &=& C_F \left( 3/2+2\ln\d'_s \right)  \nonumber\\
A_0^{q \rightarrow qg} &=& C_F \left[ 7/2 - \p^2/3 - \ln^2\d'_s - \ln\d_c \left(3/2+2\ln\d'_s \right) \right]  \nonumber\\
A_1^{g \rightarrow q\overline{q}} &=& -n_f/3  \nonumber\\
A_0^{g \rightarrow q\overline{q}} &=& n_f/3 \left( \ln\d_c-5/3 \right) \, ,
\eea
and 
\be
\d'_s = \frac{s_{12}}{s_{12}+s_{45}-M^2_{\Upsilon}} \simeq \frac{\hat{s}}{\hat{s}-1} \d_s \, .
\ee
Thus the total cross section for real correction processes in hard final state
collinear region can be written as:
\bea
\s^{HC}_f&=&\sum\limits_{i,j,k_1,k_2}\int  \hat{\s}^{HC}_f[i+j\rightarrow \uo +k_1 +k_2] \NO\\
&&\times G_{i/p}(x_1,\m_f)G_{j/p}(x_2,\m_f)\mathrm{d}x_1\mathrm{d}x_2  \NO\\
&=&\sum\limits_{i,j,k}\int  \hat{\s}^B[i+j\rightarrow \uo +k]B^{HC}(k) \NO\\
&&\times G_{i/p}(x_1,\m_f)G_{j/p}(x_2,\m_f)\mathrm{d}x_1\mathrm{d}x_2  ,
\eea
where 
\bea
B^{HC}(g)&=&\left[ \frac{\a_s}{2\pi} \frac{\Gamma(1-\e)}{\Gamma(1-2\e)} \left( \frac{4\pi\mu_r^2}{s_{12}} \right)^\e\right]
\\&&
\times\biggl( \frac{A_1^{g \rightarrow gg}+A_1^{g \rightarrow q\overline{q}}}{\e} + A_0^{g \rightarrow gg}+A_0^{g \rightarrow q\overline{q}} \biggr) \, ,\NO\\
B^{HC}(q)&=&\left[ \frac{\a_s}{2\pi} \frac{\Gamma(1-\e)}{\Gamma(1-2\e)} \left( \frac{4\pi\mu_r^2}{s_{12}} \right)^\e\right]
\biggl( \frac{A_1^{q \rightarrow qg}}{\e} + A_0^{q \rightarrow qg} \biggr) \, .\NO
\eea
\begin{table}[htbp]
\begin{center}
\begin{tabular}{|c|c|c|}
\hline\hline
$\mathrm{R}_i$&$\mathrm{L}_i^\prime$&$A^{HC}_{i}$\\
\hline
$gg\rightarrow  \uo gg$&$gg\rightarrow  \uo g$ &$\dfrac{1}{\e}A_1^{g\rightarrow gg}+A_0^{g\rightarrow gg}$ \\
\hline
$gq\rightarrow  \uo gq$ &$gq\rightarrow  \uo q$ &$\dfrac{1}{\e}A_1^{q\rightarrow qg}+A_0^{q\rightarrow qg}$\\
\hline
$gg\rightarrow  \uo q\bar{q}$ &$gg\rightarrow  \uo g$&$\dfrac{1}{\e}A_1^{g\rightarrow q\bar{q}}+A_0^{g\rightarrow q\bar{q}}$ \\
\hline
$q\bar{q}\rightarrow  \uo gg$  &$q\bar{q}\rightarrow  \uo g$&$\dfrac{1}{\e}A_1^{g\rightarrow gg}+A_0^{g\rightarrow gg}$\\
\hline
$q\bar{q}\rightarrow  \uo q\bar{q}$ &$q\bar{q}\rightarrow  \uo g$ &$\dfrac{1}{n_f}\left(\dfrac{1}{\e}A_1^{g\rightarrow q\bar{q}}+A_0^{g\rightarrow q\bar{q}}\right)$\\
\hline
$q\bar{q}\rightarrow  \uo q'\bar{q}'$ &$q\bar{q}\rightarrow  \uo g$ &$\left(1-\dfrac{1}{n_f}\right)\left(\dfrac{1}{\e}A_1^{g\rightarrow q\bar{q}}+A_0^{g\rightarrow q\bar{q}}\right)$\\
\hline\hline
\end{tabular}
\caption{The hard final state collinear factors for real correction processes and the corresponding LO processes.}
\label{table:final_coll}
\end{center}
\end{table}
\paragraph{initial state collinear}
Almost all real processes, except process (R6), contain hard initial state collinear singularities. These singularities are partly absorbed into the redefinition of the parton distribution
function (PDF) of the concerned hadrons (usually it is called as the
mass factorization \cite{Altarelli:1979ub}). Here we adopt the scale
dependent PDF using the $\overline{\rm MS}$ convention given in
Ref~\cite{Harris:2001sx}. 
\bea
G_{b/p}(x,\mu_f)&=&G_{b/p}(x)-\frac{1}{\e} \left[
\frac{\a_s}{2\pi} \frac{\Gamma(1-\e)}{\Gamma(1-2\e)}
\left(\frac{4\pi \mu_r^2}{\mu_f^2}\right)^{\e}\right] \NO\\&&\times
\int_x^1 \frac{\md z}{z} P_{bb'}(z)G_{b'/p}(x/z) \, . \label{eqn:PDF}
\eea
The second term is sometimes referred as the mass factorization counter-term. There is still something remaining after the cancellation, which can be expressed in two terms. The first one, which only exists in the real processes with final state gluon, can be expressed as
\bea
\hat{\s}^{HC}_{i}[\mathrm{R}_i]&=&\hat{\s}^B[\mathrm{L}_i]\left[ \frac{\a_s}{2\pi} \frac{\Gamma(1-\e)}{\Gamma(1-2\e)} \left( \frac{4\pi\mu_r^2}{\m_f} \right)^\e \right]A^{SC}_i , \NO
\eea
with 
\bea
A^{SC}_1&=& 2A^{SC}(g\rightarrow gg) \NO\\
A^{SC}_2&=& A^{SC}(q\rightarrow qg)+A^{SC}(g\rightarrow gg) \NO\\
A^{SC}_3&=& 2A^{SC}(q\rightarrow qg) ,
\eea
and
\bea
A^{SC}(q\rightarrow qg) &=& \dfrac{1}{\e}C_F\left[3/2+2\ln(\d_s) \right] \nonumber\\ 
A^{SC}(g\rightarrow gg) &=&  \dfrac{1}{\e}\left[2N \ln \d_s + (11N-2 n_f)/6\right]  \,.
\eea
The corresponding hadronic total cross section is 
\bea
\s^{HC}_i&=&\sum\limits_{i,j,k}\int  \hat{\s}^{HC}_i[i+j\rightarrow \uo +k+g] \NO\\
&&\times G_{i/p}(x_1,\m_f)G_{j/p}(x_2,\m_f)\mathrm{d}x_1\mathrm{d}x_2  \NO\\
&=&\sum\limits_{i,j,k}\int  \hat{\s}^B[i+j\rightarrow \uo +k][B^{SC}(i)+B^{SC}(j)] \NO\\
&&\times G_{i/p}(x_1,\m_f)G_{j/p}(x_2,\m_f)\mathrm{d}x_1\mathrm{d}x_2  ,
\eea
with
\bea
B^{SC}(g)&=&\left[ \frac{\a_s}{2\pi} \frac{\Gamma(1-\e)}{\Gamma(1-2\e)} \left( \frac{4\pi\mu_r^2}{\m_f} \right)^\e \right] A^{SC}(g\rightarrow gg) \NO\\
B^{SC}(q)&=&\left[ \frac{\a_s}{2\pi} \frac{\Gamma(1-\e)}{\Gamma(1-2\e)} \left( \frac{4\pi\mu_r^2}{\m_f} \right)^\e \right] A^{SC}(q\rightarrow qg) .\NO
\eea
The other term is obtained by summing up the remaining contributions from all the real correction processes. It can be written as 
\bea
&&\s^{HC}_{add}\left[pp\rightarrow \uo+X\right]\\
&\equiv&\sum\limits_{i,j,k}\int \hat\s^B\left[ij\rightarrow \uo+k\right] \left[
\frac{\a_s}{2\pi} \frac{\Gamma(1-\e)}{\Gamma(1-2\e)} \left(\frac{4
\pi \mu_r^2}{s_{12}}\right)^{\e}\right] \NO
\\&&\times
 \biggl[G_{i/p}(x_1,\mu_f)\widetilde{G}_{j/p}(x_2,\mu_f) +(x_1\leftrightarrow x_2)\biggr] \md x_1 \md x_2, \NO
\eea
with 
\be \widetilde{G}_{c/p}(x,\mu_f) =
\sum_{c'}  \int_x^{1-\d_s\d_{cc'}} \frac{dy}{y} G_{c'/p}(x/y,\mu_f)
\widetilde{P}_{cc'}(y) \, , \label{eqn:g_tilde} \ee 
and 
\be
\widetilde{P}_{ij}(y) = P_{ij}(y)\ln\left(\d_c\frac{1-y}{y}
\frac{s_{12}}{\mu_f^2}\right) - P_{ij}^{\prime}(y) \, . \ee
The $n$-dimensional unregulated ($y<1$) splitting functions $P_{ij}(y,\e)$ has been written as $P_{ij}(y,\e)=P_{ij}(y)+\e P_{ij}^\prime(y)$ with
\bea
P_{qq}(y) &=& C_F \frac{1+y^2}{1-y} ,\nonumber \\
P_{qq}^{\prime}(y) &=& -C_F(1-y) ,\nonumber\\
P_{gq}(y) &=& C_F \frac{1+(1-y)^2}{y} ,\nonumber \\
P_{qq}^{\prime}(y) &=& -C_Fy ,\nonumber\\
P_{gg}(y) &=& 2N\left[ \frac{y}{1-y}+\frac{1-y}{y}+y(1-y)\right] ,\nonumber\\
P_{gg}^{\prime}(y) &=& 0 ,\nonumber\\
P_{qg}(y) &=& \frac{1}{2} \left[ y^2+(1-y)^2 \right] ,\nonumber\\
P_{qg}^{\prime}(y) &=& -y(1-y) \, .
\eea
\subsection{Cross section of all NLO contributions}
The hard noncollinear part $\s^{H\overline{C}}$ is IR finite and can be numerically computed using the standard Monte-Carlo integration techniques. Now the real cross section can be expressed as
\be
\s^R=\s^S+\s^{HC}_f+\s^{HC}_i+\s^{HC}_{add}+\s^{H\bar{C}}.
\ee
And we have
\be
\s^{NLO} =\s^B+\s^V+\s^R.
\ee

\section{Transverse momentum distribution}
To obtain the transverse momentum $p_t$ distribution of $\Upsilon$, a similar
transformation for integration variables ($\md x_2 \md t \rightarrow
J\md p_t \md y$) which we introduced in our previous work \cite{Gong:2008hk} is applied.  Therefore we have \bea \dfrac{\md \s}{\md
p_t}= \sum_{i,j} \int J \md x_1 \md y G_{i/p}(x_1,\mu_f)G_{j/p}(x_2,\mu_f)
\dfrac{\md \hat \s}{\md t}, \eea 
with
\bea
&p_1=x_1\dfrac{\sqrt{S}}{2}(1,0,0,1),
&p_2=x_2\dfrac{\sqrt{S}}{2}(1,0,0,-1), \NO\\[3mm]
&m_t=\sqrt{M_{\Upsilon}^2+p_t^2},
&p_3=(m_t \cosh y,p_t,0,m_t \sinh y),\NO\\[3mm]
&x_t=\dfrac{2m_t}{\sqrt{S}},
&\tau=\dfrac{m_4^2-M_{\Upsilon}^2}{\sqrt{S}},\\[3mm]
&J=\dfrac{4 x_1 x_2 p_t}{2x_1-x_t e^y},
&x_2=\dfrac{2\tau+x_1~x_t e^{-y}}{2 x_1-x_t e^y}, \NO\\[3mm]
&x_1|_{min}=\dfrac{2 \tau + x_t e^y}{2- x_t e^{-y}},&\NO
\eea
where $\sqrt{S}$ is the center-of-mass energy of $p\bar{p}(p)$ at Tevatron or LHC,
$m_4$ is the invariant mass of all the final state particles except $\Upsilon$, and
$y$ and $p_t$ are the rapidity and transverse momentum of $\Upsilon$ in the laboratory frame respectively.
\section{Polarization}
The polarization parameter $\alpha$ is defined as: 
\be
\alpha(p_t)=\frac{{\md\s_T}/{\md p_t}-2 {\md\s_L}/{\md p_t}}
                 {{\md\s_T}/{\md p_t}+2 {\md\s_L}/{\md p_t}}.
\ee 
It represents the measurement of $\Upsilon$ polarization as
function of $\Upsilon$ transverse momentum $p_t$ when calculated at
each point in $p_t$ distribution. To evaluate $\alpha(p_t)$, the polarization of $\Upsilon$ must
be explicitly retained in the calculation. The partonic differential
cross section for a polarized $\Upsilon$ is expressed as: \be
\dfrac{\md \hat{\s}_{\lambda}}{\md t}= a~\epsilon(\lambda) \cdot
\epsilon^*(\lambda) + \sum_{i,j=1,2} a_{ij} ~p_i \cdot
\epsilon(\lambda) ~p_j \cdot \epsilon^*(\lambda), \label{eqn:polar}
\ee where $\lambda=T_1,T_2,L$.
$\epsilon(T_1),~\epsilon(T_2),~\epsilon(L)$ are the two transverse
and longitudinal polarization vectors  of $\Upsilon$ respectively, and
the polarizations of all the other particles are summed over in
n-dimension. One can find that $a$ and $a_{ij}$ are finite when the
virtual corrections and real corrections are properly handled as
aforementioned. Therefore there is no difference
in the differential cross section ${\md \hat{\s}_{\lambda}}/{\md t}$
whether the polarization of $\Upsilon$ is summed over in 4 or $n$
dimensions. Thus we can just treat the polarization vectors of
$\Upsilon$ in 4-dimension, and also the spin average factor goes back
to 4-dimension. The gauge invariance is explicitly checked by
replacing the gluon polarization vector into its 4-momentum in the
final numerical calculation. 

\section{Treatment of $\jpsi$}
The production mechanism of $\jpsi$ at Tevatron and LHC is much
similar to that of $\Upsilon$ except that, color-octet states
contribute much more in $\jpsi$ production according to the
experimental data and LO theoretical predictions. The results of above calculation can also be applied to the case of $\jpsi$ by doing the
substitutions: \bea
m_b&\leftrightarrow& m_c   \NO\\
M_{\Upsilon}&\leftrightarrow& M_\jpsi \NO\\
R_s(0)^{\Upsilon}&\leftrightarrow& R_s(0)^\jpsi \\
n_f=4 &\leftrightarrow& n_f=3  \NO \eea 
Note that in $\jpsi$ production, charm quark is no longer treated as light quark.
\section{numerical result}
In our numerical computations, the CTEQ6L1 and CTEQ6M
PDFs \cite{cteq}, and the corresponding fitted value $\a_s(M_Z)=0.130$ and
$\a_s(M_Z)=0.118$  are used for LO and NLO calculations respectively.
The bottom quark mass is set as $4.75 \gev$. 

The choice of the renormalization scale $\mu_r$ and factorization scale $\mu_f$ is an important issue in the calculations, and it causes uncertainties. We choose $\mu=\mu_r=\mu_f=\sqrt{(2m_b)^2+p_t^2}$ as our default choice.
And the center-of-mass energies are chosen as 1.96 TeV at Tevatron and 14 TeV at LHC.

At First, different values of the two cutoffs, $\d_s$ and $\d_c$, are used to to check the independence of the final results on the cutoffs and the invariance is observed within the error tolerance. Then the two phase space cutoffs are fixed as $\d_s=10^{-3}$ and $\d_c=\d_s/50$ in the following calculations. 

It is known that the QCD perturbative expansion is not good in the regions
of small transverse momentum or large rapidity of $\Upsilon$. Therefore, the results are
restricted in the region $p_t>3$. For the rapidity cut,  $|y_{\Upsilon}|<1.8$ is chosen at the Tevatron,
the same cut condition as the experiments~\cite{Abazov:2005yc}, and at the LHC, it is chosen to be $|y|<3$.

\begin{figure}
\center{
\includegraphics*[scale=0.45]{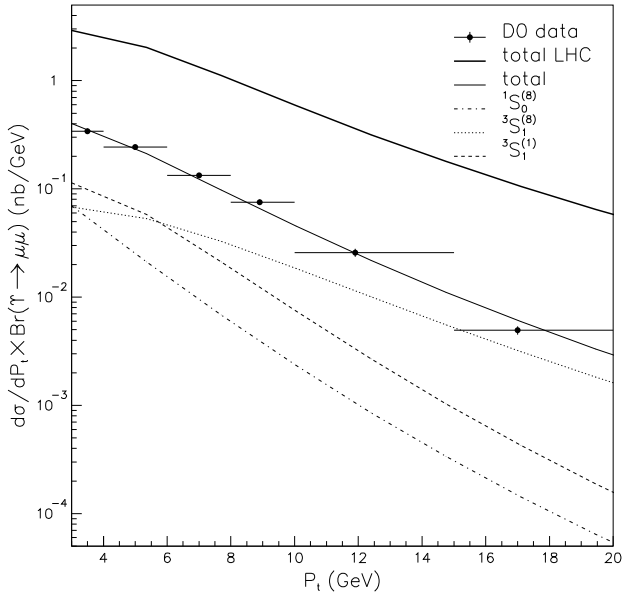}
\caption {\label{fig:fit_inc}Transverse momentum distribution of inclusive $\Upsilon$ production at Tevatron and LHC.
The D0 data is from Ref.~\cite{Abazov:2005yc}.}}
\end{figure}

To fix the NRQCD matrix elements for color-octet states of $\Upsilon(1S)$, the D0 data~\cite{Abazov:2005yc} is used, and the fitting starts from Eq.(4) of Ref.~\cite{Braaten:2000cm} where the contributions from spin-singlet states $\eta_b(nS)$ and $h_b(nS)$ are not included. And we have to take a few approximations in our fitting procedure:

\begin{itemize}
\item For the S-wave color-singlet part, only the direct color-singlet $\Upsilon(1S)$ and feed-down from $\Upsilon(2S)$ are considered, while other contributions have been neglected. The contribution from the feed-down of $\Upsilon(2S)$ can be included to the direct $\Upsilon(1S)$ production by multiplying a factor of $Br[\Upsilon(2S)\rightarrow\Upsilon(1S)+X]\times\ME{\Upsilon}{1}{(2S)}/\ME{\Upsilon}{1}{(1S)}$, which results in a factor of 1.127 after a short calculation with PDG data~\cite{Amsler:2008zzb}. And the results for direct $\Upsilon(1S)$ of color-singlet contribution are extracted from our previous work~\cite{Gong:2008hk}.

\item The contributions from P-wave color-singlet states $\chi_{bJ}(nP)$ are estimated by multiplying a decay fraction
$F^{\Upsilon(1S)}_{\chi_b(nP)}\approx F^{\Upsilon(1S)}_{\chi_b(1P)}+F^{\Upsilon(1S)}_{\chi_b(2P)}$, where $F^{\Upsilon(1S)}_{\chi_b(1P)}$ and $F^{\Upsilon(1S)}_{\chi_b(2P)}$ can be obtained from an older sample with the cuts 
$p_t>8$ and $|y_{\Upsilon}|<0.4$~\cite{Affolder:1999wm}. As pointed out in Ref.~\cite{Artoisenet:2008fc}, the fraction should not depend very strongly on $p_t$ according to Fig.2 of Ref.~\cite{Acosta:2001gv}. Also, from Fig.4 of Ref.~\cite{Abazov:2005yc} we can see it should not depend very strongly on the rapidity cut either. Thus $F^{\Upsilon(1S)}_{\chi_b(1P)}=27.1\pm 6.9\pm4.4\%$ and $F^{\Upsilon(1S)}_{\chi_b(2P)}=10.5\pm4.4\pm1.4\%$ are taken in our calculation, which result in $F^{\Upsilon(1S)}_{\chi_b(nP)}\approx37.6\pm9.4\%$.
\item The contribution from P-wave color-octet states $\Upsilon[\pj]$ at NLO are still not available. As shown below, the NLO QCD corrections to $\jpsioa$ don't change the cross section very much. If we assume that the NLO QCD corrections to $\Upsilon[\pj]$ are also small, we can mix it with $\jpsioa$ again, like what we have done at LO. Thus, the value of our fitted $\mopa_{\mathrm{inc}}$ includes the contributions from $\Upsilon[\pj]$ as well.
\end{itemize}
With these approximations, the formula we used for the fitting of inclusive color matrix elements becomes
\bea
d\s[\Upsilon]_{\mathrm{inc}}&=&1.127\times d\s[(b\bar{b})_1(\OP{3}S{1})]\ME{\Upsilon}{1}{(\OP{3}S{1})} \NO\\
&+&F^{\Upsilon(1S)}_{\chi_b(nP)}d\s[\Upsilon]_{\mathrm{inc}} 
+d\s[(b\bar{b})_8(\OP{1}S{0})]\ME{\Upsilon}{8}{(\OP{1}S{0})}_{\mathrm{inc}} \NO\\
&+&d\s[(b\bar{b})_8(\OP{3}S{1})]\ME{\Upsilon}{8}{(\OP{3}S{1})}_{\mathrm{inc}} ,
\label{eqn:fit}
\eea
and the NRQCD matrix elements for color-octet states $\ME{\Upsilon}{8}{}_{\mathrm{inc}}$ are determined as
\bea
&&\mopa_{\mathrm{inc}}=(0.948\pm0.444)\times 10^{-2} \gev^3 \NO\\
&&\mopb_{\mathrm{inc}}=(4.834\pm0.719)\times 10^{-2} \gev^3,
\eea
where only the uncertainty in $F^{\Upsilon(1S)}_{\chi_b(nP)}$ has been considered.
The fitting is shown in Fig.~\ref{fig:fit_inc}, together with our prediction for inclusive $\Upsilon$ production at the LHC.

\begin{figure}
\center{
\includegraphics*[scale=0.45]{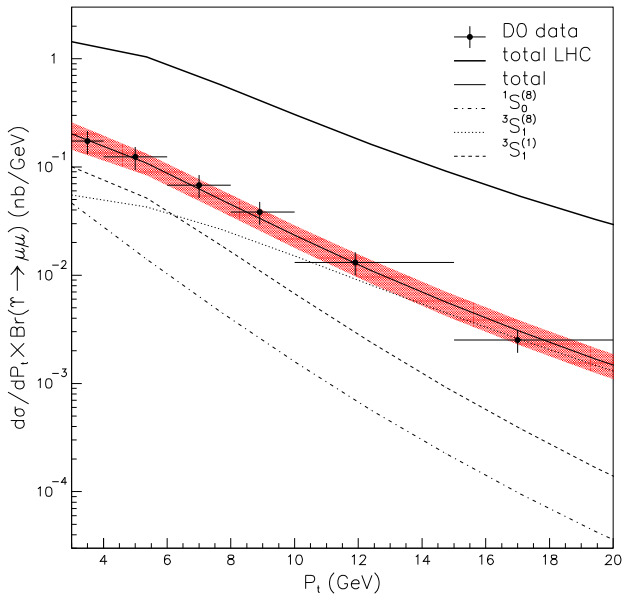}
\caption {\label{fig:fit_direct}Transverse momentum distribution of direct $\Upsilon$ production at Tevatron and LHC.
The D0 data is from Ref.~\cite{Abazov:2005yc}.}}
\end{figure}

The direct fraction of direct $\Upsilon$ production can also obtained from Ref.~\cite{Affolder:1999wm} as $F^{\Upsilon(1S)}_{\mathrm{dir}}=50.9\pm 12.2 \%$. Thus we can use the formula
\bea
F^{\Upsilon(1S)}_{\mathrm{dir}}d\s[\Upsilon]_{\mathrm{inc}}&=&d\s[(b\bar{b})_1(\OP{3}S{1})]\ME{\Upsilon}{1}{(\OP{3}S{1})} \NO\\
&+&d\s[(b\bar{b})_8(\OP{1}S{0})]\ME{\Upsilon}{8}{(\OP{1}S{0})} \NO\\
&+&d\s[(b\bar{b})_8(\OP{3}S{1})]\ME{\Upsilon}{8}{(\OP{3}S{1})} ,
\label{eqn:fit2}
\eea
to fit the direct color-octet matrix elements. The matrix elements are obtained as 
\bea
&&\mopa=(0.630\pm0.576)\times 10^{-2} \gev^3 \NO\\
&&\mopb=(3.900\pm1.063)\times 10^{-2} \gev^3,
\eea
where the uncertainty comes only from $F^{\Upsilon(1S)}_{\mathrm{dir}}$. Again the value of our fitted $\mopa$ includes the contribution from $\Upsilon[\pj]$. This fitting is shown in Fig.~\ref{fig:fit_direct}, together with our prediction for direct $\Upsilon$ production at the LHC. The band in the figure is obtained from the uncertainty of $F^{\Upsilon(1S)}_{\mathrm{dir}}$.

\begin{figure}
\center{
\includegraphics*[scale=0.45]{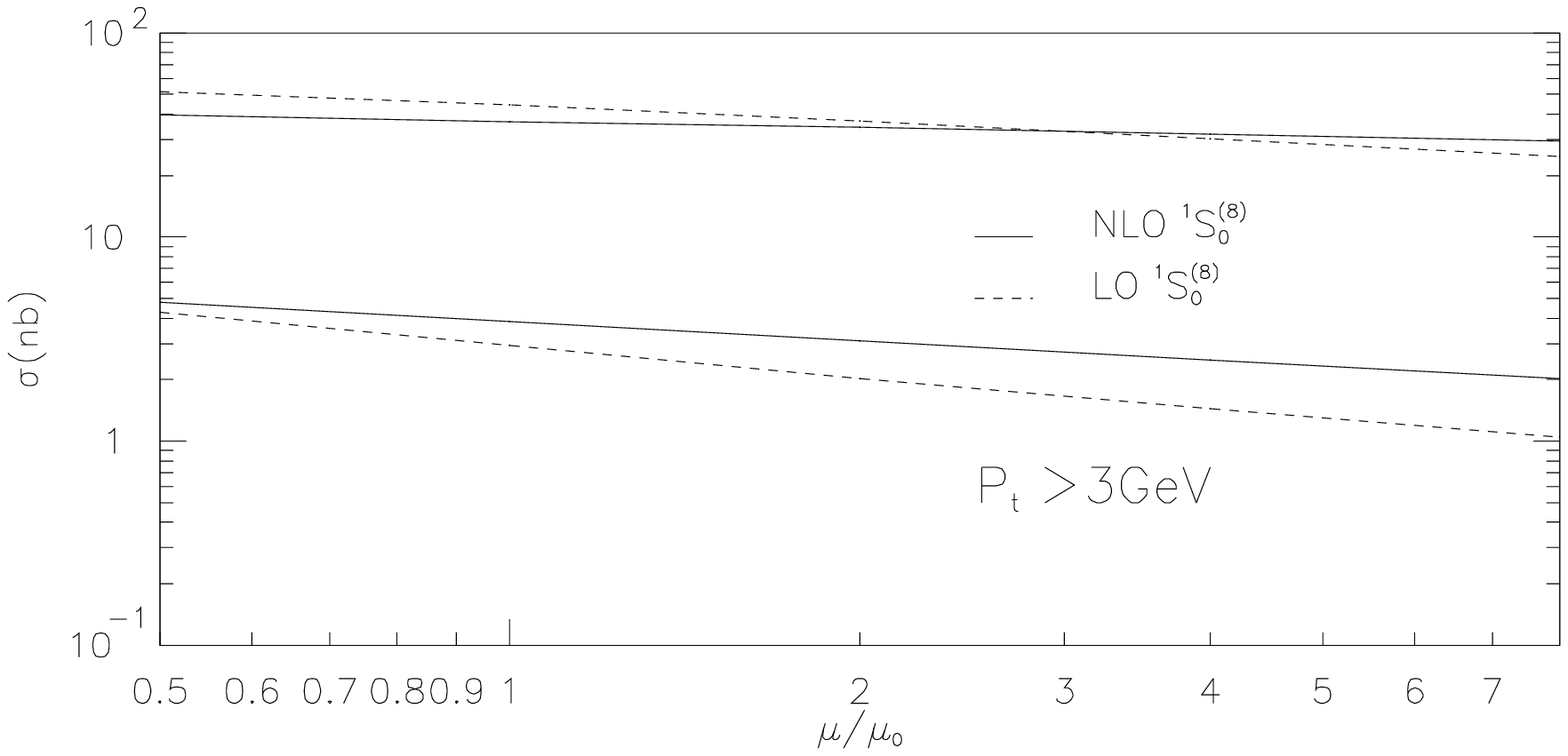}
\\
\includegraphics*[scale=0.45]{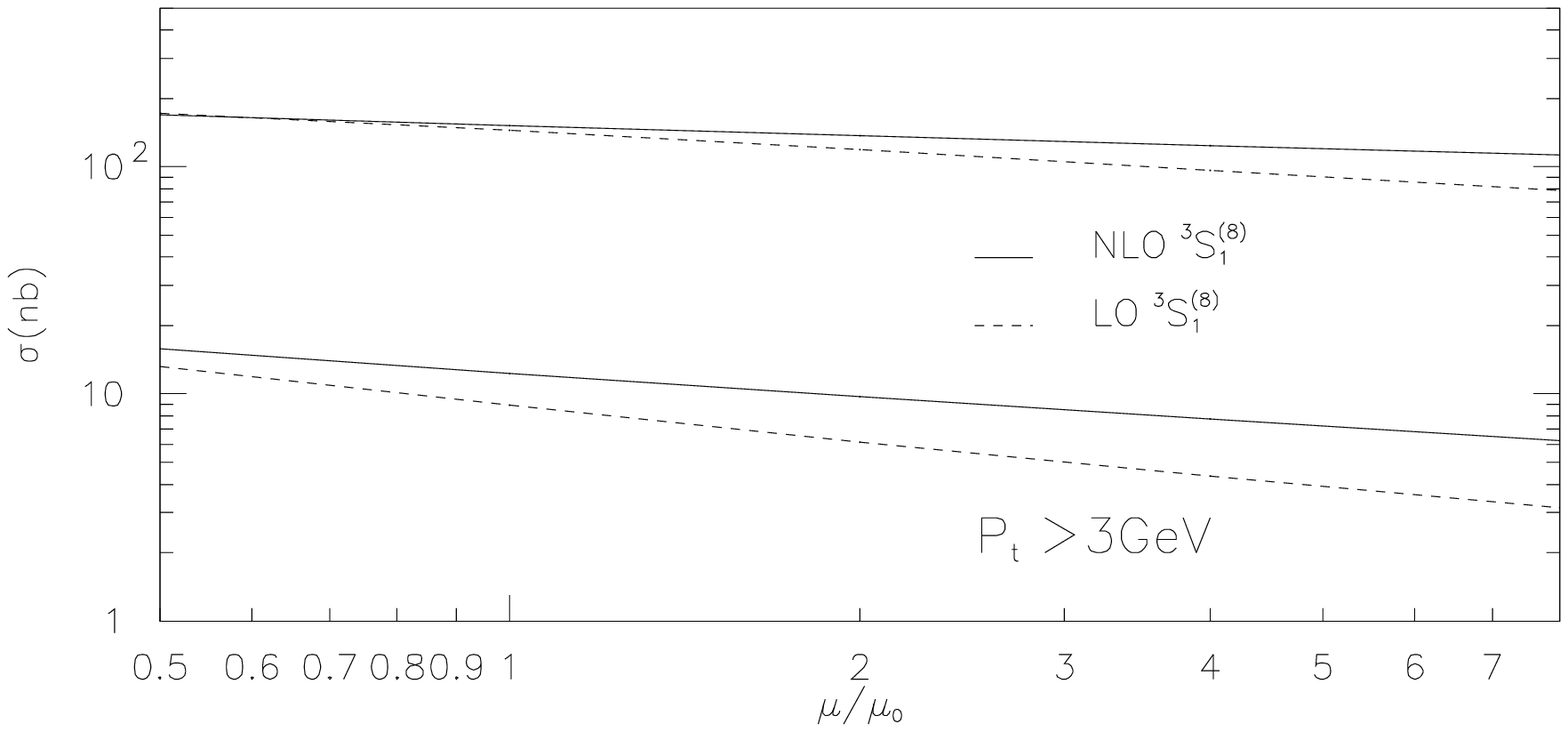}
\caption {\label{fig:total}Total cross section of $\Upsilon$ hadroproduction at LHC (upper curves) and Tevatron (lower curves), as function of $\mu$ with $\mu_r=\mu_f=\mu$ and $\mu_0=\sqrt{(2m_b)^2+p_t^2}$. 
}}
\end{figure}

\begin{figure}
\center{
\includegraphics*[scale=0.45]{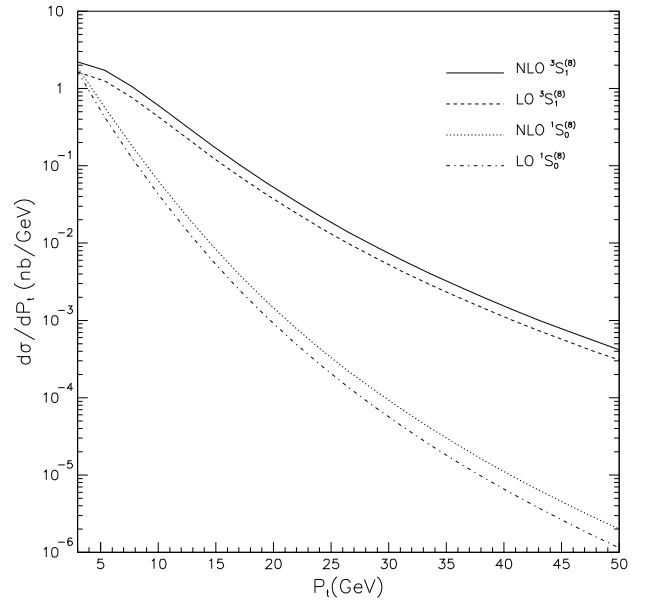}
\caption {\label{fig:pt_t}Transverse momentum distribution of $\Upsilon$ production
with $\mu_r=\mu_f=\mu_0$ at the Tevatron.}}
\end{figure}

\begin{figure}
\center{
\includegraphics*[scale=0.45]{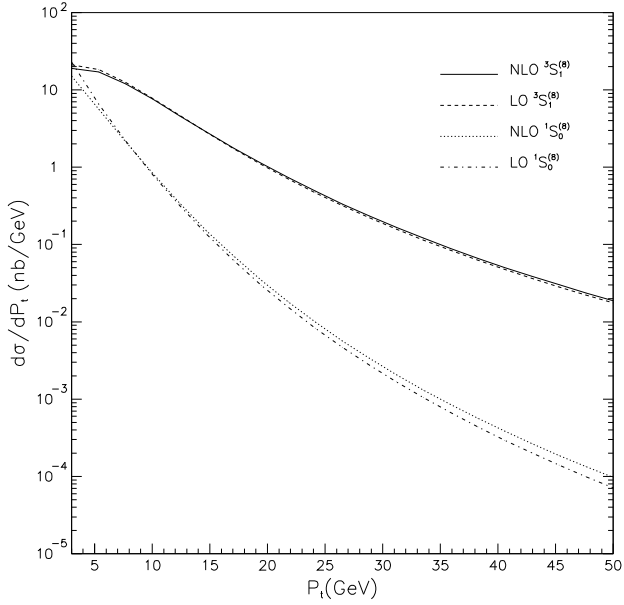}
\caption {\label{fig:pt_l}Transverse momentum distribution of $\Upsilon$ production
with $\mu_r=\mu_f=\mu_0$ at the LHC.}}
\end{figure}

The dependence of the total cross section on the renormalization
scale $\mu_r$ and factorization scale $\mu_f$ are shown in
Fig.~\ref{fig:total}. It is obvious that the NLO QCD corrections
make such dependence milder. We can also see that the NLO QCD corrections effect the cross section lesser at the LHC than at the Tevatron.
The $p_t$ distributions of $\Upsilon$ production via S-wave color-octet states are presented in Figs.~\ref{fig:pt_t} and ~\ref{fig:pt_l}, where only slight changes appear when the NLO QCD corrections are included.

\begin{figure}
\center{
\includegraphics*[scale=0.45]{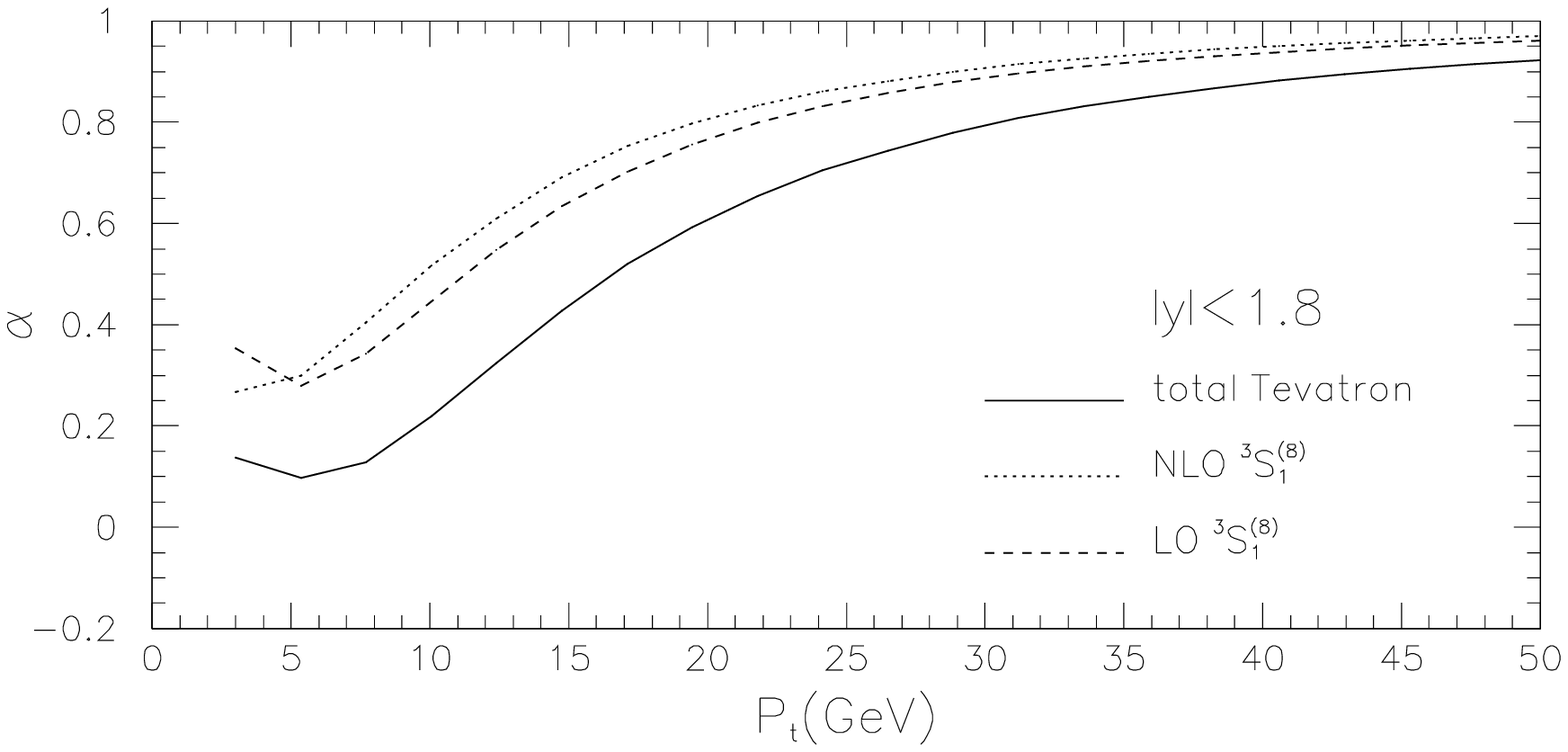}\\
\includegraphics*[scale=0.45]{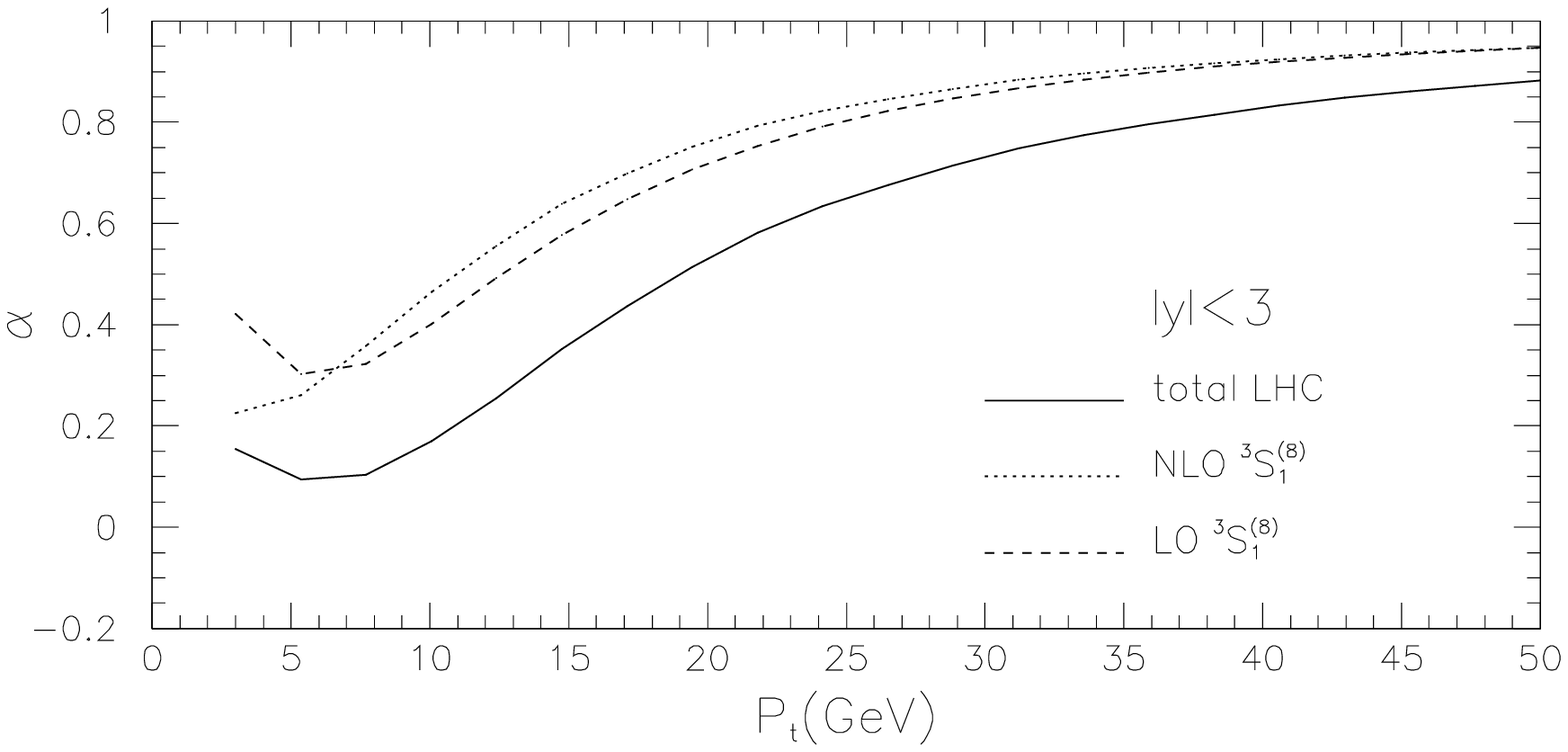}
\caption {\label{fig:polar_dir}Transverse momentum distribution of polarization parameter
$\a$ for direct $\Upsilon$ production at the Tevatron (upper) and LHC (lower).}}
\end{figure}

\begin{figure}
\center{
\includegraphics*[scale=0.45]{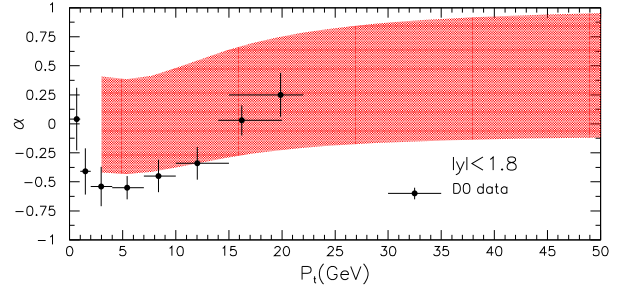}
\caption {\label{fig:polar_inc}Transverse momentum distribution of polarization parameter
$\a$ for inclusive $\Upsilon$ production at the Tevatron. The D0 data is from ref~\cite{Abazov:2008za}.}}
\end{figure}

$\jpsioa$ produces unpolarized $\Upsilon$, so it contributes to
$\a=0$ for both LO and NLO. The $p_t$ distributions of $\Upsilon$
polarization parameter $\a$ from $\jpsiob$ are shown in
Fig.~\ref{fig:polar_dir} and there is slight change when the NLO
corrections are taken into account. Our predictions for the polarization of direct $\Upsilon$ production are also presented in the figure as a "total" result. In Fig.~\ref{fig:polar_inc}, the polarization of inclusive $\Upsilon$ production at the Tevatron is shown. As the polarization of $\Upsilon$ from the feed-down of $\chi_b(nP)$ is not available yet, a huge band is obtained by verifying the polarization of this part between -1 to 1. The experimental data from the D0 is also shown in the same figure. We can see that, there is still some distance between the theoretical prediction and experimental measurement, even with such a large band.

\section{Summary and Discussion}
As a summary, in this work, we have calculated the NLO QCD
corrections to $\Upsilon$ production via S-wave color-octet states $\jpsio$ at the
Tevatron and LHC. With $\mu_r=\mu_f=\mu_0$, the K factors of total
cross section (ratio of NLO to LO) are 1.313 and 1.379 for $\jpsioa$
and $\jpsiob$ at Tevatron, while at LHC they are 1.044 and 1.182
respectively. Unlike for the color-singlet case, there are only
slight changes to the transverse momentum distributions of $\Upsilon$
production and the $\Upsilon$ polarization when the NLO QCD corrections
are taken into account. All the results imply that the perturbative
QCD expansion quickly converges for $\Upsilon$ production via the
S-wave color-octet states, in contrast with that via color-singlet,
where the NLO contributions are too large to hint a good convergence at
the NNLO. 
By fitting the experimental data from the D0 at the Tevatron, the matrix elements for S-wave color-octet states are obtained. And new predictions for the $p_t$ distributions of the $\Upsilon$ production and polarization at the Tevatron and LHC are presented. The prediction for the polarization of inclusive $\Upsilon$ contains large uncertainty rising from the polarization of $\Upsilon$ from feed-down of $\chi_b$. Even with such a large uncertainty, there are still some distance between the prediction and experiment data. 
Also, the errors of fractions used in the fitting, $F^{\Upsilon(1S)}_{\chi_b(1P)}$, $F^{\Upsilon(1S)}_{\chi_b(2P)}$ and $F^{\Upsilon(1S)}_{\mathrm{dir}}$, are quite large and result large uncertainty in the matrix elements. New measurements on the production and also polarization for direct $\Upsilon$ are expected. This would make the matrix elements more precise, and get rid of the large uncertainty from $\chi_b$.

This work is supported by the National Natural Science Foundation of
China (No.~10475083, 10979056 and 10935012), by the Chinese Academy of Science under
Project No. INFO-115-B01, and by the China Postdoctoral Science foundation
No.~20090460535.

\appendix
\section{Calculation of the factor $I^{jk}$}\label{chapter:I_jk}
If we write the $n$-momentum of soft gluon in the $p_1+p_2$ rest frame as 
\be
p_5=E_5(1, \ldots, \sin\q_1 \cos\q_2, \cos\q_1) \, ,
\label{eqn:p5_soft}
\ee
then $I^{jk}$ is defined as 
\be
I^{jk} = \int \frac{-(p_j\cdot p_k)}{(p_j\cdot p_5)(p_k\cdot p_5)} dS \, ,
\label{eqn:soft_int}
\ee
with
\bea
dS &=&  \frac{1}{\pi} \left(\dfrac{4}{s_{12}}\right)^{-\e} \int_0^{\d_s\sqrt{s_{12}}/2} 
     dE_5 E_5^{1-2\e}\sin^{1-2\e}\!\q_1\,d\q_1 
     \NO\\&&\times
     \sin^{-2\e}\!\q_2\,d\q_2 \, .\\\NO
\label{eqn:soft_PS}
\eea
Before the calculation of $I^{jk}$, define $\b_j$ as $\beta_j=|\vec{p}_j|/E_j$, which is the ratio of momentum to energy of particle $i$ in the $p_1+p_2$ rest frame, where 
\be 
\b_1=\b_2=\b_4=1,\qquad \b_3=\dfrac{\hat{s}-1}{\hat{s}+1}\equiv \b .
\ee
Then we can write $p_j$ and $p_k$ as 
\bea
p_j&=&E_j(1,\cdots,\beta_j) \NO\\
p_k&=&E_k(1,\cdots,\beta_k\sin\theta_{jk},\beta_k\cos\theta_{jk}),
\eea
where $\theta_{jk}$ is the angel between $j$ and $k$. Now we have 
\be
I^{jk} = -\frac{1-\b_j\b_k\cos\!\q}{\pi} I_E I_A^{jk} \, ,
\ee
where
\bea
I_E
&=& \left(\frac{4}{s_{12}}\right)^{-\e} \int_0^{\d_s\sqrt{s_{12}}/2}dE_5 E_5^{-1-2\e}  \nonumber\\[3mm]
&=& \left( -\frac{1}{2\e} \right) (\d_s)^{-2\e}    \, ,
\eea
and 
\bea
I_A^{jk} 
& = & \int_0^{\p} \sin^{1-2\e}\!\q_1\,d\q_1 \int_0^{\p} \sin^{-2\e}\!\q_2\,d\q_2 \dfrac{1}{1-\b_j\cos\!\q_1}
\nonumber\\ [3mm]
&&    \times \dfrac{1}{1-\b_k\cos\!\q\cos\!\q_1-\b_k\sin\!\q\sin\!\q_1\cos\!\q_2} \, .
\eea
The way to calculate the integrals $I_A^{jk}$ can be found in the appendix of Ref.~\cite{Harris:2001sx}.
Now we come to the results. It's easily to obtain 
\be
I^{11}=I^{22}=I^{44}=0,
\ee
and the others are listed below. 
\begin{enumerate}
\item {$I^{1i}~\mathrm{with}~i=2,3,4$}. Write the momenta of the particles as
\bea
p_1&=&E_1(1,\cdots,1), \NO\\
p_2&=&E_2(1,\cdots,-1), \NO\\
p_3&=&E_3(1,\cdots,\b\sin\theta_{13},\beta\cos\theta_{13}),  \NO\\
p_4&=&E_4(1,\cdots,-\sin\theta_{13},-\cos\theta_{13}), 
\eea
then we have
\bea
I_A^{12}&=&-\dfrac{\pi}{\e}, \NO\\
I_A^{13}&=&\dfrac{\pi}{1-\b\cos\theta_{13}} \biggl\{-\dfrac{1}{\e} +\ln\dfrac{(1-\hat{t})^2}{\hat{s}} 
-\e\biggl[\ln^2(1-\hat{t})
\NO\\&&
-\frac{1}{2}\ln^2\hat{s} +2\li(\hat{t}) -2\li\biggl(\dfrac{\hat{u}}{1-\hat{t}}\biggr)\biggr]\biggr\},\\
I_A^{14}&=&-\dfrac{2\p}{(1+\cos\theta_{13})\e}\biggl(\dfrac{\hat{u}}{1-\hat{s}}\biggr)^{-\e}\biggl[1+\e^2\li\biggl(\dfrac{\hat{t}}{1-\hat{s}}\biggr)\biggr], \NO
\eea
which lead to
\bea
I^{12}&=&-\dfrac{2}{\p}I_EI_A^{12}=-\dfrac{1}{\e^2}\d_s^{-2\e} \NO\\
I^{13}&=&-\dfrac{1-\b\cos\theta_{13}}{\pi}I_EI_A^{13} \NO\\
      &=& -\dfrac{1}{2\e^2}\d_s^{-2\e}  \biggl\{1 -\e\ln\dfrac{(1-\hat{t})^2}{\hat{s}} 
+\e^2\biggl[\ln^2(1-\hat{t})
\NO\\&&
-\frac{1}{2}\ln^2\hat{s} +2\li(\hat{t}) -2\li\biggl(\dfrac{\hat{u}}{1-\hat{t}}\biggr)\biggr]\biggr\},\NO\\
I^{14}&=&-\dfrac{1+\cos\theta_{13}}{\p}I_EI_A^{14} \\
      &=&-\dfrac{1}{\e^2}\d_s^{-2\e}
\biggl(\dfrac{\hat{u}}{1-\hat{s}}\biggr)^{-\e}\biggl[1+\e^2\li\biggl(\dfrac{\hat{t}}{1-\hat{s}}\biggr)\biggr]. \NO
\eea
\item{$I^{2i}~\mathrm{with}~i=3,4$}. These two can be directly obtained from $I_{1i}$ with the substitution $\hat{t} \leftrightarrow \hat{u}$.
\bea
I^{23}&=& -\dfrac{1}{2\e^2}\d_s^{-2\e}  \biggl\{1 -\e\ln\dfrac{(1-\hat{u})^2}{\hat{s}} 
+\e^2\biggl[\ln^2(1-\hat{u}) 
\NO\\&&
-\frac{1}{2}\ln^2\hat{s} +2\li(\hat{u}) -2\li\biggl(\dfrac{\hat{t}}{1-\hat{u}}\biggr)\biggr]\biggr\},\NO\\
I^{24}&=&-\dfrac{1}{\e^2}\d_s^{-2\e}
\biggl(\dfrac{\hat{t}}{1-\hat{s}}\biggr)^{-\e}\biggl[1+\e^2\li\biggl(\dfrac{\hat{u}}{1-\hat{s}}\biggr)\biggr]. \NO
\eea
\item {$I^{33}~\mathrm{and}~I^{34}$}. Write the momenta of the final state particles as 
\bea 
p_3&=&E_3(1,\cdots,-\b), \NO\\
p_4&=&E_4(1,\cdots,1),
\eea
then
\bea
I_A^{33}&=&\dfrac{2\pi}{1-\b^2}\biggl[1+\e\dfrac{1}{\b}\ln\hat{s}\biggr] ,\NO\\
I_A^{34}&=&\dfrac{\pi}{1+\b}\biggl\{-\dfrac{1}{\e}+\ln\hat{s}-\e\biggl[\frac{1}{2}\ln^2\hat{s}+2\li(1-\hat{s})\biggr]\biggr\} ,\NO\\
I^{33}&=&-\dfrac{1-\b^2}{\pi}I_EI_A^{33}=\dfrac{1}{\e}\d_s^{-2\e}\biggl[1+\e\dfrac{1}{\b}\ln\hat{s}\biggr] ,  \NO\\
I^{34}&=&-\dfrac{1+\b}{\p}I_EI_A^{34} \\
     &=&-\dfrac{1}{2\e^2}\d_s^{-2\e} \biggl\{1 -\e\ln\hat{s} +\e^2\biggl[\frac{1}{2}\ln^2\hat{s}+2\li(1-\hat{s})\biggr]\biggr\} .\NO
\eea
\end{enumerate}
\section{Color Factors}\label{chapter:color_factors}
Here we present color factors for all the processes involved. Color indices for particle $n$ are labeled as $j_n$. 
\subsection{LO processes}\label{chapter:color_factors_lo}
The color factors listed here for LO processes have been orthogonalized and normalized.
\begin{itemize}
\item $gg\rightarrow \jpsioa g$, three color factors in total:
\bea
&\dfrac{1}{\sqrt{5}}\mathrm{Tr}\bigl[T^{j_4}T^{j_2}T^{j_1}T^{j_3}- T^{j_4}T^{j_3}T^{j_1}T^{j_2}\bigr],   \NO\\
&\dfrac{1}{\sqrt{5}}\mathrm{Tr}\bigl[T^{j_4}T^{j_3}T^{j_2}T^{j_1}- T^{j_4}T^{j_1}T^{j_2}T^{j_3}\bigr],    \NO\\
&\dfrac{1}{\sqrt{5}}\mathrm{Tr}\bigl[T^{j_4}T^{j_2}T^{j_3}T^{j_1}- T^{j_4}T^{j_1}T^{j_3}T^{j_2}\bigr].
\eea
\item $gg\rightarrow \jpsiob g$, three color factors also:
\bea
&&\dfrac{1}{3\sqrt{2}}\mathrm{Tr}\bigl[
(T^{j_4}T^{j_2}T^{j_3}T^{j_1}+T^{j_4}T^{j_1}T^{j_3}T^{j_2})
\NO\\ &&~~~~~~~
-(T^{j_4}T^{j_1}T^{j_2}T^{j_3}+T^{j_4}T^{j_3}T^{j_2}T^{j_1})\bigr], \NO\\
&&\dfrac{1}{3\sqrt{6}}\mathrm{Tr}\bigl[
 (T^{j_4}T^{j_2}T^{j_3}T^{j_1}+T^{j_4}T^{j_1}T^{j_3}T^{j_2})
\NO\\&&~~~~~~~
+(T^{j_4}T^{j_1}T^{j_2}T^{j_3}+T^{j_4}T^{j_3}T^{j_2}T^{j_1})
\NO\\[2mm]&&~~~~~~~
-2(T^{j_4}T^{j_2}T^{j_1}T^{j_3}+T^{j_4}T^{j_3}T^{j_1}T^{j_2})
\bigr], \NO\\
&&\dfrac{1}{\sqrt{15}}\mathrm{Tr}\bigl[
 (T^{j_4}T^{j_2}T^{j_3}T^{j_1}+T^{j_4}T^{j_1}T^{j_3}T^{j_2})
\NO\\&&~~~~~~~
+(T^{j_4}T^{j_1}T^{j_2}T^{j_3}+T^{j_4}T^{j_3}T^{j_2}T^{j_1})
\NO\\[2mm]&&~~~~~~~
+(T^{j_4}T^{j_2}T^{j_1}T^{j_3}+T^{j_4}T^{j_3}T^{j_1}T^{j_2})
\bigr]. 
\eea
\item $gq\rightarrow \jpsioa q$, only one color factor:
\be 
\dfrac{1}{2\sqrt{15}}\bigl[
 3(T^{j_1}T^{j_3}+T^{j_3}T^{j_1})_{j_4j_2}
-\d_{j_4j_2}\d_{j_1j_3}\bigl].
\ee
\item $gq\rightarrow \jpsiob q$, two color factors:
\be 
\dfrac{\sqrt{3}}{4}(T^{j_3}T^{j_1})_{j_4j_2}, -\dfrac{1}{4\sqrt{21}}(8T^{j_1}T^{j_3}+T^{j_3}T^{j_1})_{j_4j_2}. \ee
\item $q\bar{q}\rightarrow \uo g$, almost same with $gq\rightarrow \uo q$.
\end{itemize}
\subsection{Virtual correction processes}
In the amplitude of virtual correction processes, besides the same color factors as in the corresponding LO process, there are extra ones. As we have mentioned before, virtual correction to the cross section is related to virtual amplitude as Eq.(\ref{eqn:virtual_sme}). Then the terms in proportion to these extra color factors will vanish and do not contribute to the final result as we have orthogonalized the color factors of LO processes. Thus no new color factors in virtual correction processes need to be presented here.
\subsection{Real correction processes}
In order to present the color factors of real correction processes in a simplified form, we list here all independent color factors in a certain process. Actually in our calculation, they are orthogonalized and normalized too, which are too complicated to be listed here.
\begin{itemize}
\item $gg\rightarrow \jpsioa gg$, twelve color factors. The permutations of $j_1,j_2,j_3$ and $j_4$ contain 24 terms. Divide them into twelve groups with two terms in each group, and the twelve color factors can be expressed as
\be 
\mathrm{Tr}\bigl[T^{j_5}(T^{a}T^{b}T^{c}T^{d}+T^{d}T^{c}T^{b}T^{a})\bigr],
\ee
where $a,b,c,d$ are permutations of $j_1,j_2,j_3$ and $j_4$.
\item $gg\rightarrow \jpsiob gg$, also twelve color factors. They can be expressed as
\be 
\mathrm{Tr}\bigl[T^{j_5}(T^{a}T^{b}T^{c}T^{d}-T^{d}T^{c}T^{b}T^{a})\bigr],
\ee
\item $gg\rightarrow \jpsioa q\bar{q}$, five independent color factors:
\bea
&&d^{j_2j_3k}(T^{j_1}T^{k})_{j_4j_5}, \quad d^{j_2j_3k}(T^{k}T^{j_1})_{j_4j_5}, \NO\\
&&d^{j_1j_3k}(T^{j_2}T^{k})_{j_4j_5}, \quad d^{j_1j_3k}(T^{k}T^{j_2})_{j_4j_5}, \\
&&6(T^{j_3}T^{j_2}T^{j_1}-T^{j_1}T^{j_2}T^{j_3})_{j_4j_5} 
+if^{j_1j_2j_3}\d_{j_4j_5}. \NO
\eea
\item $gg\rightarrow \jpsiob q\bar{q}$, seven independent color factors. One is $d^{j_1j_2j_3}\d_{j_4j_5}$ while the others can be expressed as
\be
(T^aT^bT^c)_{j_4j_5},
\ee
where $a,b,c$ are permutations of $j_1,j_2$ and $j_3$.
\item $gq\rightarrow \uo gq$ and $q\bar{q}\rightarrow \uo gg$, similar to $gg\rightarrow \uo q\bar{q}$.
\item $q\bar{q}\rightarrow \jpsioa q\bar{q}$, two color factors:
\be
T^{j_3}_{j_2j_5}\d_{j_4j_1}+T^{j_3}_{j_4j_1}\d_{j_2j_5}, \quad
T^{j_3}_{j_2j_1}\d_{j_4j_5}+T^{j_3}_{j_4j_5}\d_{j_2j_1}. 
\ee
\item $q\bar{q}\rightarrow \jpsiob q\bar{q}$, four color factors:
\be 
T^{j_3}_{j_2j_5}\d_{j_4j_1},~T^{j_3}_{j_4j_1}\d_{j_2j_5}, ~T^{j_3}_{j_2j_1}\d_{j_4j_5}, ~T^{j_3}_{j_4j_5}\d_{j_2j_1}. 
\ee
\item $qq\rightarrow \uo qq$, similar to $q\bar{q}\rightarrow \uo q\bar{q}$.
\item $q\bar{q}\rightarrow \jpsioa q'\bar{q}'$, only one color factor:
\be
3(T^{j_3}_{j_2j_5}\d_{j_4j_1}+T^{j_3}_{j_4j_1}\d_{j_2j_5})-2(T^{j_3}_{j_2j_1}\d_{j_4j_5}+T^{j_3}_{j_4j_5}\d_{j_2j_1}). 
\ee
\item $q\bar{q}\rightarrow \jpsiob q'\bar{q}'$, three color factor:
\bea
&&T^{j_3}_{j_2j_5}\d_{j_4j_1}-T^{j_3}_{j_4j_1}\d_{j_2j_5}, \NO\\
&&T^{j_3}_{j_2j_1}\d_{j_4j_5}-T^{j_3}_{j_4j_5}\d_{j_2j_1}, \NO\\
&&3T^{j_3}_{j_4j_1}\d_{j_2j_5}-T^{j_3}_{j_4j_5}\d_{j_2j_1}.
\eea
\item $qq'\rightarrow \uo qq'$, similar to $q\bar{q}\rightarrow \uo q'\bar{q'}$.
\end{itemize}
\bibliography{paper}
\end{document}